\documentclass[twocolumn,prl,letterpaper,groupedaddress]{revtex4}

\usepackage{times,amsmath,amsfonts,amssymb}
\usepackage{graphicx}
\usepackage{color}
\usepackage{todonotes}

\renewcommand{\phi}{\varphi}

\renewcommand{\epsilon}{\varepsilon}
\newcommand{\id}{\textbf{I}}

\newcommand{\hwp} {\mbox{HWP}}

\def\ket#1{{\lvert}#1\rangle}

\begin{document}
\title{Enhanced delegated computing using coherence}
\author{Stefanie Barz$^1$, Vedran Dunjko$^{2,3}$, Florian Schlederer$^1$, Merritt Moore$^1$, Elham Kashefi$^4$, Ian A. Walmsley$^1$}
\affiliation{$^1$~Clarendon Laboratory, Department of Physics, University of Oxford, OX1 3PU, United Kingdom,\\
$^2$~Institute for Quantum Optics and Quantum Information, Austrian Academy of Sciences, Technikerstrasse 21a, A-6020 Innsbruck, Austria\\
$^3$~Institute for Theoretical Physics, University of Innsbruck, Technikerstrasse 25, 6020 Innsbruck, Austria\\
$^4$~School of Informatics, Informatics Forum, 10 Crichton Street, Edinburgh, EH8 9AB, UK}

\begin{abstract}
A long-standing question is whether it is possible to delegate computational tasks securely. Recently, both a classical and a quantum solution to this problem were found~\cite{Gentry2009, Broadbent2009}. Here, we study the interplay of classical and quantum approaches and show how coherence can be used as a tool for secure delegated classical computation. 
We show that a client with limited computational capacity---restricted to an XOR gate---can perform universal classical computation by manipulating information carriers that may occupy superpositions of two states. Using single photonic qubits or coherent light, we experimentally implement secure delegated classical computations between an independent client and a server.
The server has access to the light sources and measurement devices, whereas the client may use only a restricted set of passive optical devices to manipulate the light beams.
Thus, our work highlights how minimal quantum and classical resources can be combined and exploited for classical computing.
\end{abstract}
\maketitle
\section{Introduction}

\begin{figure}
	\centering
		\includegraphics[width=0.33\textwidth]{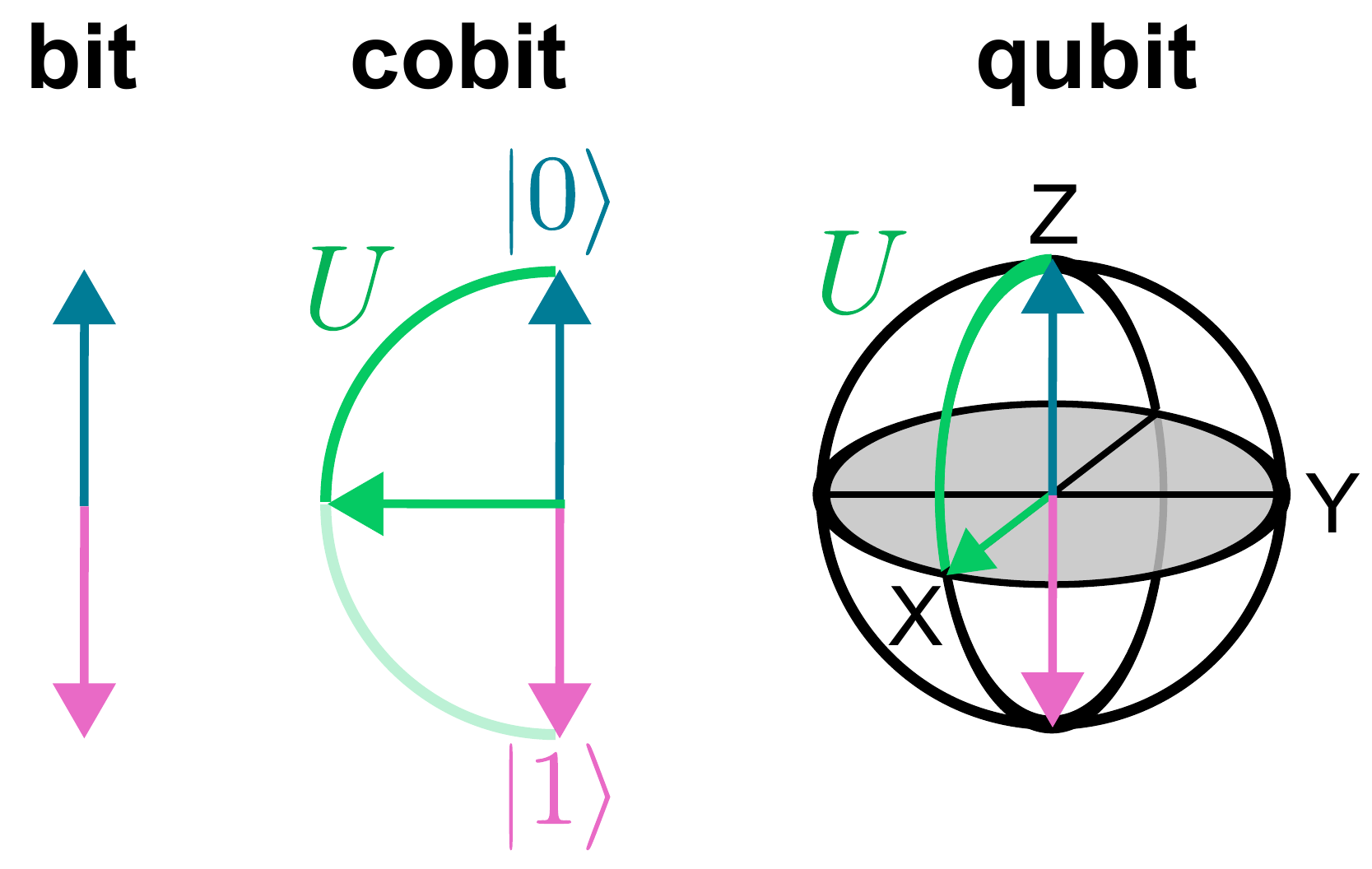}
		\caption{Bit, cobits, and qubits. The bit is a two-level classical system, cobits are systems capable of being in a coherent superposition of two "states", and qubits are quantum systems. The operation $U$ transforms basis states into superposition states and vice versa.}
	\label{fig:cobit}
\end{figure}

Cloud computing, the storage and processing of data on remote servers, has become highly relevant to modern information processing.
The question of whether it is possible to compute over encrypted data was first asked some 35 years ago~\cite{Rivest1978}. With the progress from stand-alone machines to large connected networks, the security of delegated computations has become increasingly important.
In 2009, a classical algorithm, the fully homomorphic encryption protocol, was invented which provides computation security in data processing at remote servers~\cite{Gentry2009}.
At the same time, a quantum computing protocol was found which allows an almost-classical client to delegate a quantum computation securely to a quantum server~\cite{Broadbent2009, Barz2012}. In contrast to the classical algorithm, the quantum version provides unconditional security~\cite{Broadbent2009, Morimae2012, Giovannetti2013, Fisher2014}; however, it requires classical communication of the order of the size of the computation. 
The trade-off between the amount of communication required and the desired security level is what motivates evaluation of a hybrid quantum-classical scheme~\cite{Tan2014}. 

Here, we study the interplay between classical and quantum delegated computation. 
The central question is what kind of additional resources a client, with capability restricted only to parity computations (XOR), needs in order to perform universal classical computations and to delegate those securely to a server.
We show that this can be accomplished using \textit{cobits}, systems capable of being in a coherent superposition of two "states" (see Fig.~\ref{fig:cobit}), for example single photonic qubits or coherent laser beams.

In our scheme, the server has access to cobits, and the client is restricted to parity computations and the local manipulation of the cobits. The protocol works in the following manner: the server sends cobits, and the client applies simple operations to them, dependent on some classical bits. The cobits are then sent back to the server, which performs a measurement. The result of the measurement depends on the client's manipulations and contains the encrypted outcome of the NAND operation on the client's classical bits. 
This means that the cobit enables the client to compute problems beyond her own power, since the NAND gate is universal for classical computation.

Further, we experimentally implement classical secure delegated computation by using single qubits or coherent laser beams as cobits. In our implementation, the client and the server are set up in two different laboratories, separated by more than 50 meters, and connected by optical fibres. 
Photonic systems are ideally suited for this task, since they can be easily manipulated and transmitted over large distances; however our scheme can be implemented using every physical system that provides coherence.

Note that the protocol and the implementation are classical in the sense of classical physics: they use purely classical means, effects and devices, including classical coherence. We note that this definition differs from the definition of "classical" in computer science, which is limited to only classical two-level bits and gates on these bits.
Thus, our work also highlights the two different notions of classicality in physics and computer science.

\section{theory}

\begin{figure}
	\centering
		\includegraphics[width=0.42\textwidth]{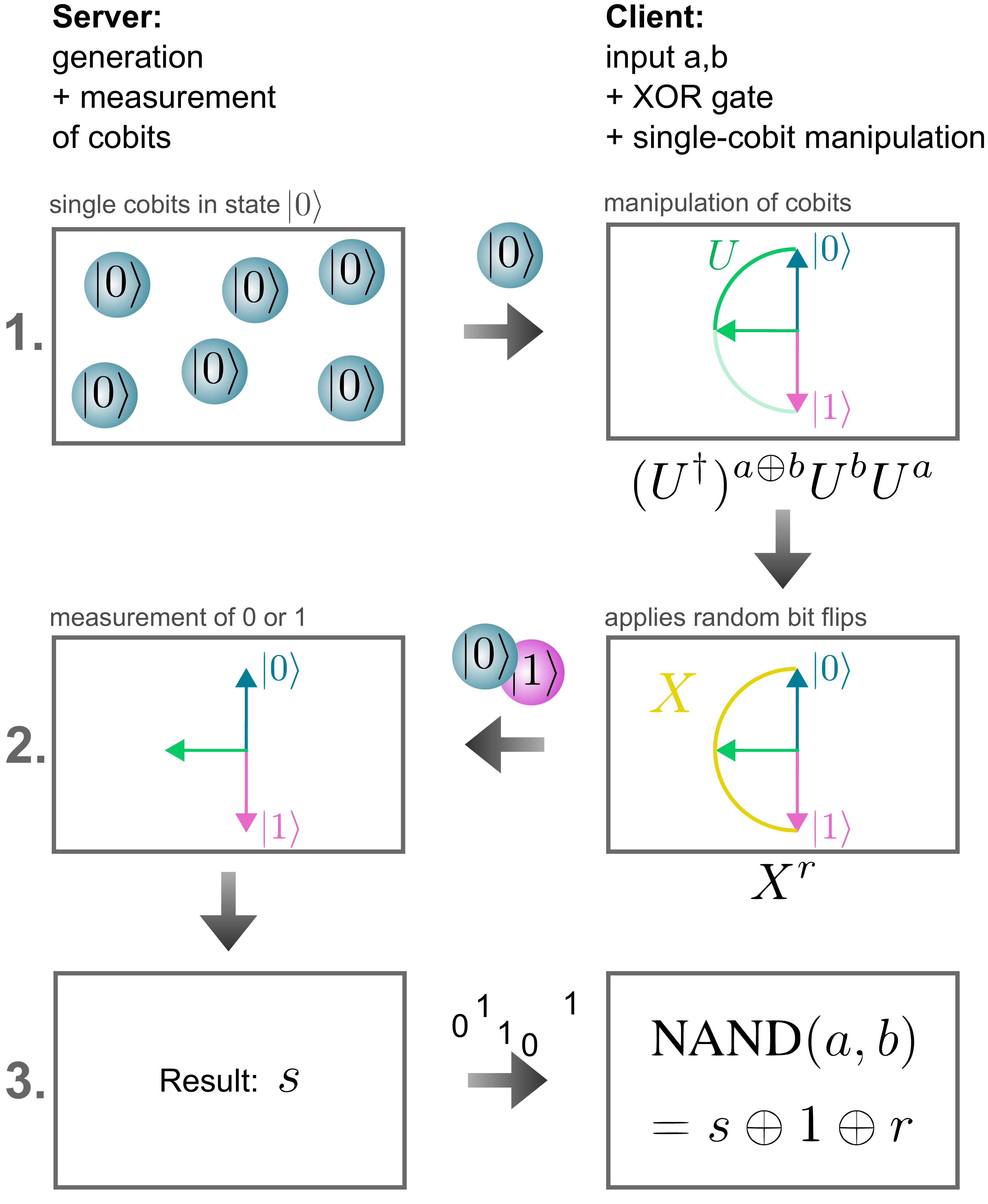}
	\caption{Scheme of delegated NAND gate. The steps of the protocol are in detail described in the main text.}
	\label{fig:scheme}
\end{figure}

Our work is based on a protocol for secure delegated classical computation using quantum resources~\cite{Dunjko2014}. 
It was shown that manipulations of only two-level bits are not sufficient for this task. Here, we reformulate the original work~\cite{Dunjko2014} and show that in the same setting adding classical coherence enables us to perform secure delegated classical computations. 

The protocol is based on the implementation of a NAND gate using only parity computations and coherence. Here, we first describe the protocol using single cobits 
and show later its implementation with single photonic qubits and coherent beams, which relaxes the requirements of the initial theory~\cite{Dunjko2014}.
In detail, the protocol works as explained in the following (see also Fig~\ref{fig:scheme}).
First, the server generates cobits in the state~$\ket{0}$ and sends these cobits to the client.
The client wants to implement a NAND gate on two input bits $a$ and $b$.
The client encodes the result of a NAND(a,b) gate in the output cobit by applying the gate sequence:
\begin{equation}
	\ket{\mbox{NAND}(a,b)\oplus 1}= (U^{\dagger})^{a\oplus b} U^b U^a\ket{0}.
\end{equation}
Here, $U$ is an operation which brings the state $\ket{0}$ into a superposition of $\ket{0}$ and $\ket{1}$. If $U$ is applied to the superposition of $\ket{0}$ and $\ket{1}$, the cobit will be in state $\ket{1}$ after the operation ($U(U\ket{0})=\ket{1})$.
In our protocol, the operation $U$ is or is not applied, depending on the values of $a$ and $b$.
Only if $a=b=1$, the output cobit is in state $\ket{1}$, for all other settings of a and b, the output cobit is in state $\ket{0}$. Thus, the output cobit can be written as $\ket{\mbox{NAND}(a,b)\oplus 1}$ and effectively contains a NAND gate.

In order to hide the state of the output cobit to achieve secure delegated computing, 
 the client applies an additional random bit flip X:
\begin{equation}
	\ket{\mbox{NAND}(a,b)\oplus 1 \oplus r}= X^r\ket{\mbox{NAND}(a,b)\oplus 1},
\end{equation}
where r is a random value.

The cobit is then sent back to the server, where a measurement in the $\ket{0/1}$ basis is performed.
The result of this measurement, s, is returned to the client, who finally obtains the result NAND(a,b) by computing:
\begin{equation}
	\mbox{NAND}(a,b)=s \oplus 1 \oplus r.
\end{equation}



A single classical bit is not sufficient to  implement a NAND gate, because at least two bits are required. Our protocol shows that systems allowing for a coherent superposition of two states are sufficient.
A single qubit also accomplishes this task in the fully quantum case. Here, the operation $U=R_y(\pi/2)$ is a rotation of $\pi/2$ around the Y axis of the Bloch sphere: $R_y(\theta)= \exp{(-i \theta/2 \sigma_y)}$, $\sigma_y$ is the Pauli operator, and the bit flip $X=\sigma_X$ is given by the Pauli operator .
However, no quantum behavior is required in our setting.
Every system that provides coherence can be used to implement our protocol.

Optics facilitates transmission of information between the server and the client and back.
Experimentally, we make use of single photonic qubits or a coherent laser beam, since the logical states $\ket{0}$ and $\ket{1}$ can be encoded in the photon's or beam's polarization. 
The only difference is that when using a coherent state light beam multiple photons pass through the client's gates with the same settings.
Since the security effectively reduces to a classical information-theoretical encryption (effectively a one-time pad) and is not relying on quantum properties vital in most of quantum cryptography (e.g. the no-cloning result for quantum states),
having multiple copies of the same state does not reduce the
security (see proof in SI).

The challenge when single qubits are used for the protocol is that probabilistic generation and optical losses affect the robustness of the protocol.
Since the client is only capable of performing parity computations and the preparation of random bits, she cannot check whether the computation is correct or not.
If the server does not send a photon or the photon gets lost, then the server fails to register a result.
The easiest solution would be to send an additional classical bit on a different channel from the server to the client, which indicates that the procedure has worked. Dependent on the classical bit, the client could then repeat the computation. However, this is not possible in our framework as this routine would be equivalent to implementing a NAND gate and thus is beyond the client's capabilities.
Using a laser beam for the implementation of the protocol has the advantage of providing robustness against these photon losses.

A NAND computation without considering the security aspects, was first proposed in another work~\cite{Anders2009}. There, a classical parity computer
controlled three-qubit Greenberger-Horne-Zeilinger states in order to perform universal classical computation. This setting can be seen as a measurement-based version of ours---a rotation is performed via single-qubit measurements~\cite{Raussendorf2001, Raussendorf2003}.
Our work shows that the same functionality can be achieved without having any quantum resources at all. 
Furthermore, we achieve secure delegated computations by sending cobits. This reduction to the manipulation of "simple" resources, compared to the generation of entanglement, clearly decreases the experimental requirements and enables one to perform secure and delegated classical computations with minimal resources.


\section{Experiments}

\begin{figure}
	\centering
		\includegraphics[width=0.49\textwidth]{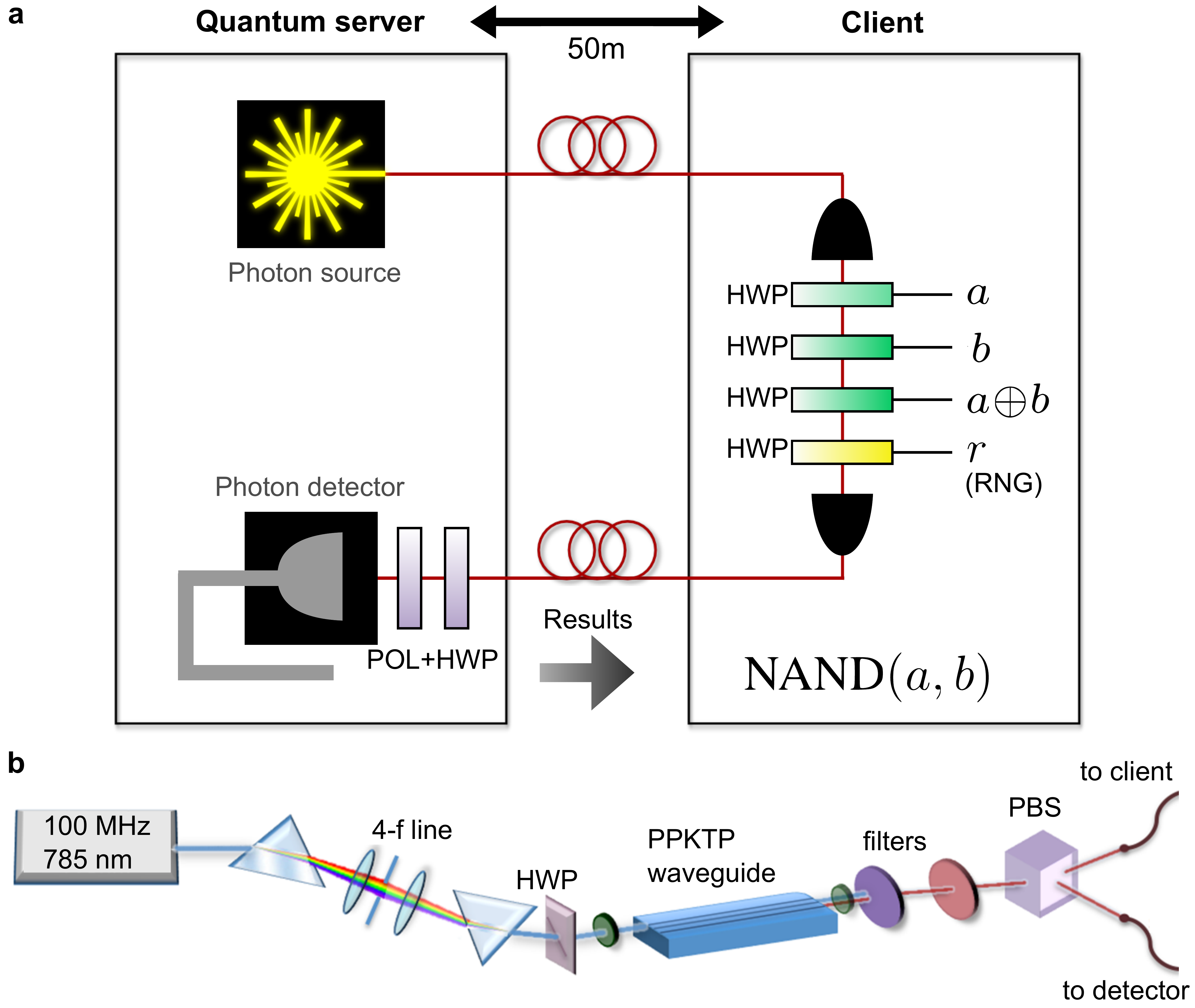}
		\caption{Experimental setup. a. Setup of separated client and server. The server in ``lab 1'' generates and measures polarization-encoded single qubits or the polarization of an attenuated laser beam. The client in ``lab 2'' manipulates the polarization and encodes the NAND gate.
		b. Source for the generation of heralded single photons. }
	\label{fig:setup}
\end{figure}
We implement the server and the client using two independent experimental setups running in two different laboratories, which are separated by 50 m (see Fig.~\ref{fig:setup}).

We either use a heralded single photon source or a weak coherent laser beam for the implementation of the protocol. For both cases, we encode the states $\ket{0}$ and $\ket{1}$ in polarization, denoting horizontal and vertical polarization, respectively.

The heralded single photons are produced by type-II parametric down conversion in a Potassium Titanium Oxide Phosphate (KTP) crystal that has periodically poled waveguides~\cite{Harder2013}. A mode-locked fiber-based femtosecond laser produces $90\,$~fs long pulses at $1575\,$~nm with a repetition rate of $100\,$~MHz. These pulses are frequency-doubled in a 1 mm long periodically poled Potassium Dihydrogen Phosphate (KDP) crystal cut for type-II second harmonic generation, resulting in $7\,$~mW of $787\,$~nm light. The fundamental 1575 nm light is filtered out with a dichroic mirror and short-pass filter, and the $787\,$~nm beam is focused through $3\,\mu$m wide waveguides in a $10\,$~mm long AR-coated KTP crystal, which is periodically poled to phase-match for type-II parametric down-conversion. After the chip, long-pass filters 
are used to block out the pump light. The horizontally and vertically polarized down-converted photons, centered at 1570 nm and 1580 nm, are split with a polarizing beam splitter cube.  
The photons are further filtered and coupled into single-mode fibers.
The photons at 1570 nm are guided to the client's setup, whereas the photons at 1580 nm are kept the server's side and produce the heralding signal. 
Alternatively, we use a coherent laser beam at $1550\,$nm that is attenuated to the single photon level.

These polarization-encoded cobits are sent to the client who implements the required gates using wave plates.
We show in the Supplementary Information (SI) that it is sufficient for the client to have access to three half-wave plates (HWP) for the implementation of the NAND gate and to one additional HWP for the implementation of the $X^r$ operation.
By applying the following gate sequence:
\begin{equation}\label{fullgatesequence2} 
\underbrace{\hwp(\phi^r)}_{\mbox{$X$ or $\id Z$}}.\underbrace{\hwp({-\theta}^{(a\oplus b)}).\hwp({\theta}^{-b}).\hwp({\theta}^{a})}_{\mbox{gate implementation}}
\end{equation}
with $\phi=\pi/4$ and $\theta=\pi/8$, the client alters the output state, dependent on the values of $a$ and $b$. The value of the random number $r$ is generated via a classical computer in our implementation. 
However, this could be easily replaced by a quantum random number generator.

The output cobit is send back to the server who performs a measurement in the computational basis. Experimentally, for both implementations, the polarization of the photons returned to the server is analyzed using a half-wave plate, a Glan-Thompson polarizer and InGaAs avalanche photo diodes 
that are specified to be $20\%$ efficient and a deadtime set to 10 $\mu$s. 
The results of the server's measurement is then equal to AND(a,b).

Note, that a real physical implementation introduces state-dependent phase shifts, for example $\hwp(0)=\sigma_z$. In order to avoid that these phase shifts reveal any information about $a$ or $b$, the settings have to be carefully chosen. As we show in the SI, the settings given above are secure in the sense, that these global phase shifts do not reveal any information about our computation.
Further, additional phase shifts are introduced when the photons are sent through the fibers. These phase shifts are independent of the settings of $a$ and $b$ and do not affect the correctness of the computation.

\section{Results}
\begin{figure}
	\centering
		\includegraphics[width=0.45\textwidth]{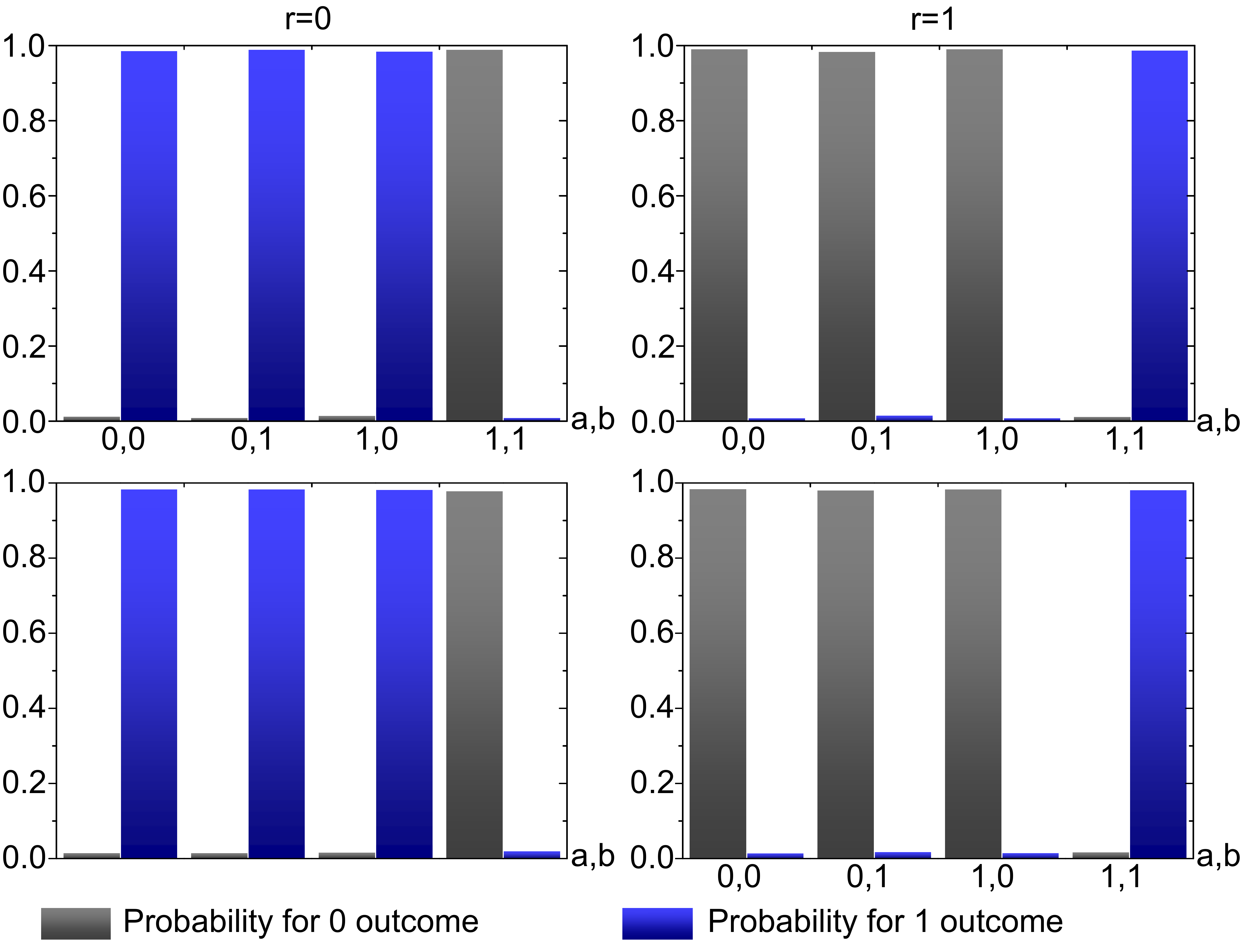}
	\caption{Results of delegated secure NAND gate. Implementation with single photons (top row) and 
	with an attenuated laser beam (bottom row) for the cases $r=0$ (left) and $r=1$ (right).  We achieve probabilities for finding the correct output of $(98.8\pm 0.5)\,\%$	for the single-photon implementation and of $(98.2\pm 0.06)\,\%$	for the implementation with a coherent beam.}	
	\label{fig:results}
\end{figure}

We first implement the protocol with single photons.
Since the protocol is secure even when multiple photons pass at the same time though the same settings (see SI), a single-shot implementation is not necessary and we integrate the result over $10$s of measurement time.
In our experiment, we use a Glan-Thompson polarizer and an additional HWP for analysing the polarization. 
The results of the single-photon runs are shown in Fig.~\ref{fig:results}a. We obtain count rates of 300 heralded photons per second. The average probability for finding the correct results is $(98.8\pm 0.5)\,\%$.

We run the same experimental sequence with a laser beam that is attenuated to 30000 single counts per second, measured after the transmission through the setup. 
In this experimental run, we obtain similar average probabilities of finding the correct results of $(98.2\pm 0.06)\,\%$ (see detailed results in Fig.~\ref{fig:results}b). 
In both experiments, the errors are calculated assuming Poissonian errors.
Experimental imperfections arise from polarizations drifts when the photons are transmitted through fibers and errors in the manipulations with wave plates as well as imperfection in the measurement in the $\ket{0,1}$ basis.

\begin{figure}
	\centering
		\includegraphics[width=0.40\textwidth]{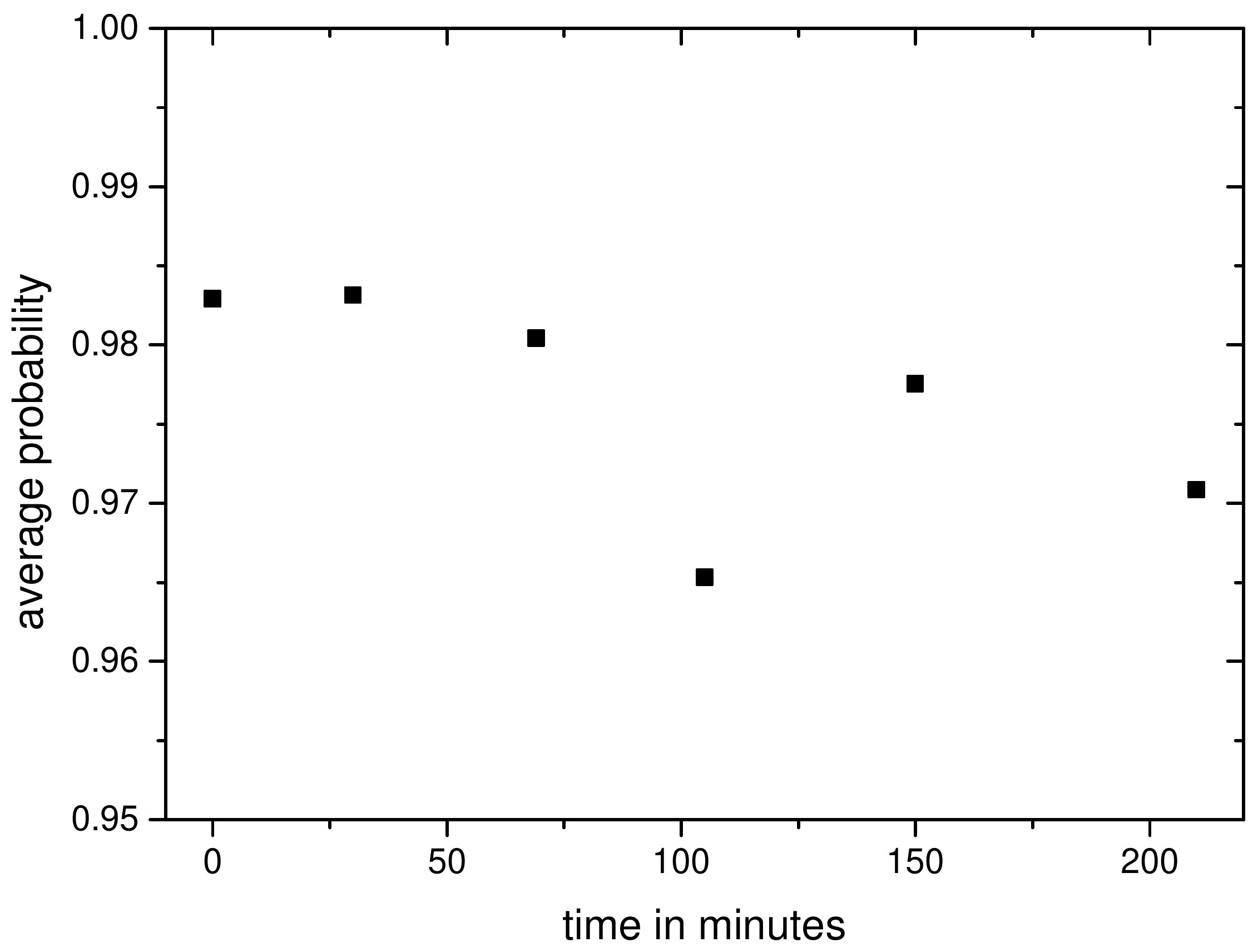}
	\caption{Study of the long-term stability of our experiment. We repeat the measurement sequence, shown in Fig.~\ref{fig:results}, six times over 210 minutes and compute the average probability of obtaining the correct result of the NAND computation (averaged over all results, for $r=0$ and $r=1$). Error bars are not shown as they are smaller than the symbols.}
	\label{fig:averageresults}
\end{figure}

The fibres connecting both laboratories are $50\,$m long and are placed partly outside the building. 
In order to test the long-term stability of our fibre connection and influences such as temperature changes and movements of the fibres, we perform a series of NAND-gate measurements for all possible inputs and repeat this measurement six times over 210 minutes. During this period, the obtained probabilities are stable and decrease only slightly from on average $(98.2\pm 0.06)\,\%$ to $(97.1\pm 0.08)\,\%$  (see Fig.~\ref{fig:averageresults}).

\section{Conclusion}
In this work, we have studied secure delegated computing at the boundary between classical and quantum physics. We have shown that the computational power of classical entity limited to parity computations can be boosted to universal classical computation by exploiting coherence.
We have shown that a single qubit can be used as a simple system to accomplish this task---even though no quantumness is required. The extension of previous work to systems capable of being in a coherent superposition of two states provides a practical and robust way to implement the protocol experimentally while still being secure.

We note that the protocol we present here is completely classical in the sense of classical physics. In a different setting, it could also be accomplished with a classical pointer instead of qubits and coherent beams. Here, the classical pointer represents a three-level system, which shows the same functionality than a two-level system with coherence. 
However, this would also require the client to have a different functionality.

While the focus of our work is more of fundamental nature, demonstrating the computational capability of cobits, a potential practical application of it could be also investigated in future. 
Note that any partial efficient classical solution for secure cloud computing once boosted to be universal would require a huge overhead. We intend to explore whether our scheme could be used as an alternative scheme where the more costly encoding for NAND computing is done via cobits. 

Furthermore, our implementation can be easily extended to long distances using standard technology from quantum key distribution.
In the future, it will be interesting to study how this scheme can be extended to multi-party computations, where different parties compute a result while hiding the inputs from each other.

In conclusion, our work shows a new way of how to exploit the properties of both quantum particles and classical fields as tools for classical computing.

\section{Acknowledgements}
We thank Animesh Datta, Andreas Eckstein, Peter Humphries, Steve Kolthammer, Ben Metcalf, and Josh Nunn for discussions.
This work was supported by the Marie Curie Actions within the Seventh Framework Programme for Research of the European Commission, under the Initial Training Network PICQUE, Grant No. 608062 and by the UK Engineering and Physical Sciences Research Council (EPSRC EP/K034480/1).


\appendix
\section{Supplementary Information}

\subsection{Correctness of the experimental implementation}
The original protocol requires gates to be applied conditioned on the values of $a$ and $b$~\cite{Dunjko2014}.
However, when using polarization and wave plates, these might apply state-dependent phase shifts.
For example, a half-wave plate (HWP) at ``0'' setting is equivalent to a $\sigma_Z$ gate, at a setting of $\pi/8$, it is a Hadamard gate, and at $\pi/4$ it is an $\sigma_X$ gate. 
In order to avoid that these state-dependent phase shifts leak information to the server, we need to choose the settings carefully and ensure that the output state contains no information about $a$ and $b$.

To this end, we choose the following sequence for the implementation of the NAND gate:
\begin{equation}\label{HWProtation}
	\hwp({-\theta}^{(a\oplus b)}).\hwp({-\theta}^{b}).\hwp({\theta}^{a})\ket{0},
\end{equation}
with $\theta=\pi/8$. 
For the settings $a=b=0$, $a=0, b=1$, $a=1, b=0$, this gate sequence adds an additional phase shift of $\pi$ to the state $\ket{1}$. This phase shift can be compensated if we incorporate an additional phase flip in our one-time pad. 
For this, we use another wave plate $\hwp(\phi^r)$ with $\phi=\pi/4$, which allows us to randomly switch between a phase flip and a bit flip.
Thus, we can implement the whole scheme using only four HWPs securely:
\begin{equation}
\hwp(\phi^r).	\hwp({-\theta}^{(a\oplus b)}).\hwp({-\theta}^{b}).\hwp({\theta}^{a})\ket{0}
\end{equation}
with $\phi=\pi/4$ and $\theta=\pi/8$.

\subsection{Security of implementation using laser beams}
The security of the implemented protocol can follow immediately from the proof given in~\cite{Dunjko2014} under two assumptions: 
\begin{enumerate}
	\item ideal devices and or devices with noise/loss, provided the noise/loss parameters are not controlled by the server.
	\item the malevolent server does send individual photon states in the modes that ensure the correct operation of the optical elements on the client's side on the polarization degrees of freedom of the photons, e.g. correct frequency of light.
\end{enumerate}
Next we show that the security is not jeopardized under a broader choice
of malevolent activity by the server, which can be straightforwardly applied to the coherent light setting. 

The cumulative action of optical devices on the client's side are easily seen to implement a polarization rotation of zero degrees,  if
$\textup{NAND}(a,b) \oplus r = 0$, and $\pi$ otherwise. 
In other words, the map itself, implemented by the client, is classically one-time padded.
Thus, irrespective of the of the actual state prepared by the server, the action of such a map results in a state that is one-time padded by the parameter
$r$, so independent from the client's inputs, when averaged over the client's secret parameter $r$. The latter means the protocol is blind.

We note that the security may be jeopardized if the server utilizes other modes, e.g. frequency of light,  which changes how the optical devices, on the side of
the client, manipulate the polarization degrees of freedom. However, such behavior can in principle be prevented by quality control, which sporadically checks the characteristics of light used by the server. More general analyses of how particular implementations may be vulnerable to
attacks are beyond the scope of this work.

\end{document}